\documentclass[prd,10pt,aps,twocolumn,showpacs,nofootinbib,floatfix,amssymb,amsmath]{revtex4-1}

\usepackage{amsmath, amsfonts, amssymb}
\usepackage{graphicx}
\usepackage{soul}
\usepackage{array}

\newcommand{\be}{\begin{equation}}
\newcommand{\ee}{\end{equation}}
\newcommand{\al}[1]{\begin{align}#1\end{align}}

\newcommand{\PP}{\mathcal{P}}
\newcommand{\Pp}{\mathcal{P}'}
\newcommand{\braket}[1]{\langle #1\rangle}

\begin{document}

\title{Properties of symmetry restoration in the electroweak vacuum in very high magnetic fields}

\begin{abstract}
		We discuss properties of the electroweak vacuum as a function of an external magnetic field. The interest in these properties arises due to possible existence of the electromagnetically superconducting phase of QCD in the background of a sufficiently strong magnetic field. 
		In the electroweak theory, a superconducting vacuum exists for a well defined range of magnetic background fields, and the interest of the current paper is the behaviour near the upper critical magnetic field. We determine this critical field and calculate the vacuum condensates to leading order. 	
\end{abstract}

\author{Jos Van Doorsselaere}
\affiliation{Department of Physics and Astronomy, Ghent University, Krijgslaan 281, S9, B-9000 Gent, Belgium}

\pacs{12.38.-t, 13.40.-f, 74.90.+n}

\date{April 16, 2013}

\maketitle

\section{Introduction}

		In the past few years, interest has been renewed into the effect of strong magnetic fields on the vacuum. It was pointed out  \cite{Chernodub:2010qx} that in the case of strong interactions, the lightest vector mesons undergo a phase transition and condense into an Abrikosov lattice \cite{Abrikosov}. The vector mesons form a new condensed state of vector mesons which  is similar to the magnetised state of the electroweak vacuum. The structure of the condensates in the electroweak vacuum was first described by Ambjorn and Olesen \cite{Ambjorn:1988tm} more that 20 years ago. Recently it was found that the main feature of both -- electroweak and rho meson -- condensed structures is their unusual -- magnetic field induced --  superconductivity \cite{Chernodub:2011,Chernodub:2012fi}.  This particular kind of superconductivity is instated above a critical value of the applied magnetic field and the condensates are reinforced with growing magnetic fields. By comparison, in ordinary superconductivity, mediated by Cooper pairs, the superconductivity breaks down above the critical magnetic field. Both models, for electroweak vector bosons and rho mesons, contain a $SU(2)$ multiplet of vector fields, and there's little wonder that the results are found to be so similar. The main distinction between QCD and the electroweak theory is the scale difference: Because the masses of the rho-mesons are about 100 times smaller than the W-mass, the critical magnetic field is $10^4$ times weaker.
		
		A first look at the superconducting regime -- involving mostly classical calculations -- has given some interesting results above but very close to the critical magnetic field $B_{c1}$. Working close to the critical field allows to work in the limit where the newly created vector condensates are very small. The thus found condensates are similar for both the superconducting QCD and electroweak vacuum. The main conclusion is that a hexagonal lattice of Abrikosov vortices appears as lowest energy state for the charged vector condensate. The neutral condensates are essentially derived form this hexagonal structure and form a more involved pattern.
		
		For the electroweak vacuum it is known that at very high magnetic fields, way above the one for which vector boson condensation and superconductivity sets in, another phase transition occurs \cite{Linde:1975gx}. The Higgs condensate will vanish at this critical magnetic field, $B_{c2}$, similar to what one observes at the electroweak phase transition at some critical temperature. Indeed, at higher magnetic fields, the full SU(2) group is a symmetry of the classical vacuum. As we will show, none of the SU(2) gauge fields will be present in the vacuum as a condensate.  The structure of condensates at such enormous magnetic fields are relevant for example in baryogenesis scenario's \cite{Grasso:2000wj}.
		
\begin{figure}[]    
			\begin{center}\includegraphics[scale=0.5]{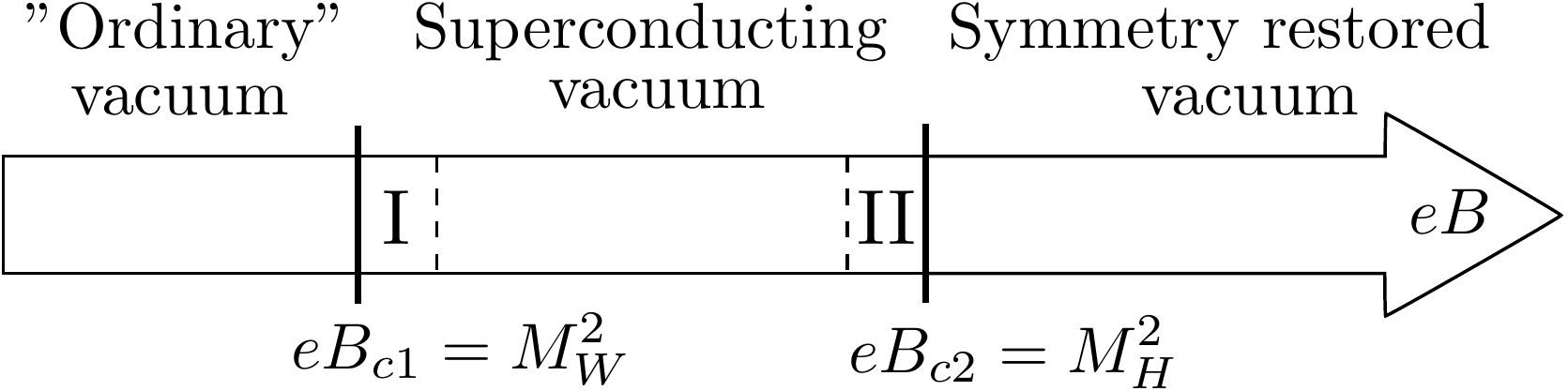} \end{center}
			\caption{The electroweak vacuum at zero temperature as a function of a constant magnetic background. While the condensates in the superconducting vacuum obey non-linear equations of motion, accurate approximations can be found for small vector condensates (region I) and small Higgs condensates (region II).}\label{diagram}
			\end{figure}

		It is the goal of the present paper to give a description of the vector condensates in magnetic fields near, this second critical magnetic field (region II in Fig.\ref{diagram}). A description was already given in the case of the non-physical Bogomolnyi limit \cite{Ambjorn:1989bd}, an analysis that can be extended to the general case, as we will show. The Higgs condensate will be approximately vanishing while the SU(2) group is still (softly) broken. This allows for condensates of chromomagnetic vortices while they are absent in the pure Yang-Mills case. We give an expression for these condensates similar (and in some cases equivalent) to the one found by Olesen \cite{Olesen:1991df} by solving a Liouville equation in terms of elliptic Weierstrass functions. 
		
		We start by giving a description of the electroweak theory in function of the relevant degrees of freedom, those transversal to the applied magnetic field. From this we find non-trivial solutions to the pure Yang-Mills equations, describing approximately the vector condensates near $B_{c2}$, the region II of Fig.\ref{diagram}. Next we explain the phase transition in detail and argue that, classically, all condensates should become trivial in the symmetry restored phase.
		
\section{The electroweak model in a magnetic field}\label{sec2}

		The bosonic electroweak Lagrangian in its simplest form separates SU(2) gauge fields, the Higgs field and the U(1) hypercharge fields. 
			\be 
			\mathcal{L}=-\frac{1}{4}W^a_{\mu\nu}W^{a\mu\nu}-\frac{1}{4}X_{\mu\nu}X^{\mu\nu}+\vert D_\mu\Phi\vert^2-V(\Phi),
			\ee
		While the latter are conventionally denoted by $B_\mu$, we have adopted the notation $X_\mu$ instead to avoid confusion with the magnetic field $B$. 
		
		A more suitable form is found when defining the electrically charged fields $W_\mu^\pm$, for which we have to fix the direction of the Higgs vacuum expectation value: $\Phi=(0,\phi)$. One finds then:
			\begin{multline}
			\mathcal{L}=-\frac{1}{2}W^+_{\mu\nu}W^{-\mu\nu}-\frac{1}{4}w^3_{\mu\nu} w^{3\mu\nu}-g w^3_{\mu\nu}W^+_\mu W^-_\nu\\-\frac{g^2}{4}(W^+_\mu W^-_\nu-W^+_\nu W^-_\mu)^2+\frac{g^2}{2}W^+_\mu W^{-\mu}\vert \phi\vert^2\\-\frac{1}{4}X_{\mu\nu}X^{\mu\nu}+\vert D_\mu\phi\vert^2-V(\phi)\label{lagrang}.
			\end{multline}
		with the notation
		\begin{align}
			W^\pm_{\mu\nu}&=\frac{1}{\sqrt{2}}(W^1_{\mu\nu}\mp iW^2_{\mu\nu})=\mathfrak{  D}_\mu W^+_\nu-\mathfrak{  D}_\nu W^\pm_\mu ,\\
			w^3_{\mu\nu}&=W^3_{\mu\nu}+ig(W^+_\mu W^-_\nu-W^+_\nu W^-_\mu),\\&=\partial_\mu W^3_\nu-\partial_\nu W^3_\mu , \\
			\mathfrak{D}_\mu W^-&=\partial_\mu W^-+igW^3_\mu W^-,\\&=\partial_\mu W^-+ig(\sin\theta A_\mu+\cos\theta Z_\mu)W^- ,\\
			D_\mu\phi&=\partial_\mu+i\frac{g}{2\cos\theta}Z_\mu\phi .
		\end{align}
		The magnetic background field will be assumed to be directed in the third spatial direction and we will refer to the (1,2)-plane as the transversal plane. It is clear that only polarisations in this plane couple to the background, and may form magnetically induced condensates. Restricting ourselves to transversal effects, it is more convenient to work with the energy density instead of the Lagrangian formalism and to use the complex coordinates $z=x_1+ix_2$ and $\bar z=x_1-ix_2$ and likewise for the fields:
			\be
			 W=W^-_1+iW^-_2,\ \bar Z=Z_1-iZ_2, \ \mathfrak{\bar{  D}}=\mathfrak{  D}_1-i\mathfrak{  D}_2\ldots
			\ee
		The vacuum equations of motion are then fully determined by the energy density as a function of transversal degrees of freedom.
			\begin{multline}
			\mathcal{ E}=
			\frac{1}{2}B^2+\frac{1}{2}\left\vert\mathfrak{\bar {  D}}W-\mathfrak{ {  D}}\bar W\right\vert^2+\frac{g^2}{2}\vert\bar W\vert^2\vert\phi\vert^2+\frac{1}{2}Z_{12}^2\\
			-\frac{1}{2}(eB+g\cos\theta Z_{12})(\vert W\vert^2-\vert \bar W\vert^2)+\frac{g^2}{8}(\vert W\vert^2-\vert\bar W\vert^2)^2\\
			+\frac{g^2}{4}(\vert W\vert^2-\vert \bar W\vert^2)\vert\phi\vert^2-\frac{1}{2}M_H^2\vert \phi\vert^2+\frac{\lambda}{4}\vert\phi\vert^4\\
			+\frac{1}{2}\left(\vert D\phi\vert^2+\vert\bar D\phi\vert^2\right)
			\label{potential}.
			\end{multline}

		An important observation in the above potential (\ref{potential}) is that only the second term depends on the phases of $W$ and $\bar W$, while the rest of the potential reaches its minimum at
			\be
			\bar W\equiv 0\label{barW},
			\ee
		for non a non-vanishing Higgs condensate. While it is not certain that one can minimise the full potential with (\ref{barW}), this seems a reasonable approximation. It is known that for small $W$ condensates, $B\gtrsim B_{c1}$, solutions assuming (\ref{barW}) exist and we will show that vacuum solutions for small Higgs condensates, $B\lesssim B_{c2}$, exist with $\bar W=0$ and with the phase of $W$ such that
			\be
			\bar{\mathfrak D}W=0.\label{abrik}
			\ee
		 Therefore we can eliminate $\bar W$ from (\ref{potential}) entirely, and with some manipulations one can rewrite it as follows:
			\begin{multline}
			\mathcal{E}=\frac{1}{2}\left(F_{12}-\frac{\cos^2\theta}{g\sin\theta}\mu_B^2-\sin\theta\frac{g}{2}\vert W\vert^2\right)^2\\
			+\frac{1}{2}\left(Z_{12}-\frac{g}{2\cos\theta}\vert\phi\vert^2+\frac{\cos\theta}{g}\mu_B^2-\cos\theta\frac{g}{2}\vert W\vert^2\right)^2\\
			+\left\vert D\phi\right\vert^2+\frac{1}{2}\left(\mu_B^2-M_H^2\right)\vert\phi\vert^2+\left(\frac{\lambda}{4}-\frac{g^2}{8\cos^2\theta}\right)\vert\phi\vert^4,\label{bb}
			\end{multline}
		where the left out boundary terms are:
			\begin{multline}
			\mathcal{E}\vert_\partial=\frac{\partial\bar\partial\vert\phi\vert^2-\bar\partial(\phi^\dag D\phi)-\partial(\phi \bar D\phi^\dag)}{2}
			\\+\mu_B^2\frac{\cos\theta F_{12}-\sin\theta Z_{12}}{g\tan\theta}-\frac{1}{2}\frac{\cos^4\theta}{g^2\sin^2\theta}\mu_B^4\label{bbound}.
			\end{multline}
		The special case for which the coefficient of the last term in (\ref{bb}) vanishes, gives an easy solution for all values of the magnetic background field, setting
			\be
			\mu_B=M_H=M_Z.
			\ee
		This Bogomolnyi limit was treated in detail in \cite{Ambjorn:1989bd}. 
		
		It is clear that for small $\vert\phi\vert$, which case is realised in the vacuum for a magnetic background field $B\lesssim B_{c2}$, the fourth order term is irrelevant and the conclusions about the phase transition at $B_{c2}$ are independent of the Bogomolnyi limit and valid for physical values of the parameters. 
				
\section{The vacuum with an infinitesimal Higgs condensate}

		Keeping Fig.\ref{diagram} in mind, we will be considering the region II close to $B_{c2}$ on the magnetic axis, where we can define a small order parameter:
			\be
			\epsilon^2=\braket{\vert\phi\vert^2}.
			\ee
		We will consider only the condensates to leading order in $\epsilon$. It is clear that this does not imply  the fourth order terms in the Higgs field, such that we are left with a sum of squares in (\ref{bb}). The first two terms can be minimised independently, giving linear equations of motion in the vacuum:
			\al{
			\sin\theta F_{12}+\cos\theta Z_{12}&=\frac{g}{2}\vert W\vert^2+\frac{g}{2}\vert\phi\vert^2,\label{w12}\\
			\cos\theta F_{12}-\sin\theta Z_{12}&=\frac{\cos\theta}{e}\mu_B^2+\cot\theta \frac{g}{2}\vert\phi\vert^2.\label{X12}
			}
		What is left is a Ginzburg-Landau like model depends on the Z-flux and Higgs field only:
			\be
			\left\vert D\phi\right\vert^2+\frac{1}{2}\left(\mu_B^2-M_H^2\right)\vert\phi\vert^2+\mu_B^2\frac{\cos\theta F_{12}-\sin\theta Z_{12}}{g\tan\theta}
			\ee
		where one of the boundary terms is now explicitly written. The decoupling of the equations of motion into linear ones by introducing $\mu_B$ only gives a nonzero leading contribution to the Higgs condensate if we take 
		\be
			\mu_B^2=M_H^2.\label{muB}
		\ee
		We are left then with the equation
			\be
			D\phi =\left(\frac{\partial}{\partial\bar z}-\frac{g}{8\cos\theta}(-2iZ)\right)\phi=0.
			\ee
		As $Z_{12}$ is a topological quantity in the presence of a Higgs condensate,  boundary conditions are independent of local fluctuations. If we solve the above equation for a homogeneous $Z_{12}=\braket{Z_{12}}$, for which we have locally $-2iZ=z\braket{Z_{12}}$, we see that physical, bounded solutions are possible only for 
			\be 
			\braket{Z_{12}}\leq 0.
			\ee
		Now we see that minimising the energy by means of the Z-flux is entirely determined by the boundary term in (\ref{bbound}) and we find
			\be
			\braket{Z_{12}}=0.\label{Z12}
			\ee
		The presence of a nonzero net flux on the contrary would have corresponded to Z-strings \cite{Nambu:1977ag} with zeroes in the Higgs condensate, organised in an Abrikosov lattice structure.  The absence of Z-strings near $B_{c1}$ was already investigated in \cite{Garriga:1995fv}. Now it becomes clear that also higher magnetic fields do not spoil this property.
		
		From (\ref{X12}), (\ref{muB}) and (\ref{Z12}) we can now identify the exact value of the second critical magnetic field by taking $\vert\phi\vert\rightarrow 0$:
			\be
			eB_{c2}=\mu_B^2=M_H^2.\label{bc2}
			\ee
		This coincides with the result of \cite{Ambjorn:1989bd} in the Bogomolnyi limit.
		
\begin{figure}[h!]
			\begin{center}
			\includegraphics[scale=0.5]{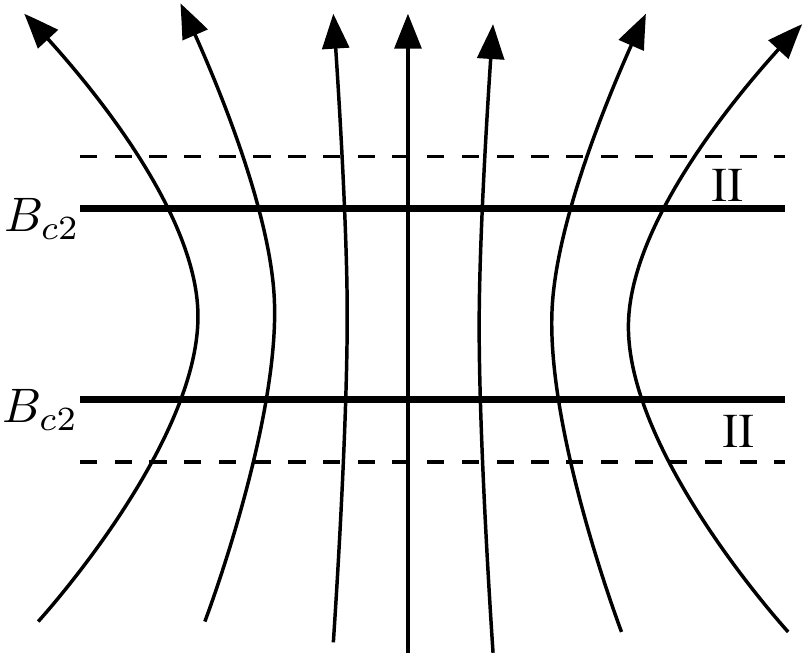}
			\caption{Squeezed magnetic flux lines, exceeding the critical value $B_{c2}$ in a compact region. In the region II, only a very small Higgs condensate is present, vanishing at the critical magnetic field. In between the full lines, the flux propagates as hyperflux only, and no condensates are present. }
			\label{fluxlines}
			\end{center}
			\end{figure}

		Moreover, one can rewrite Eq.(\ref{w12}) at $B_{c2}$
			\be
			W_{12}^3=w_{12}^3-\frac{g}{2}\vert W\vert^2=0 \label{boundary}
			\ee
		because $\phi$ vanishes there. This gives us a clear picture of what happens in a compact region of space where the magnetic field exceeds $B_{c2}$, Fig.\ref{fluxlines}. 
		At the boundary of such a region the pure Yang-Mills vacuum equations are exactly satisfied according to (\ref{boundary}), and inside --in absence of a Higgs condensate-- the equations of motion are SU(2) symmetric. Therefore one finds no more vector condensates in this region at the classical level.\footnote{Quantum effects invalidate this statement. \cite{Savvidy:1977as}}
		The magnetic flux does therefore propagate as (conserved) U(1) hypercharge flux:
			\be
			g'X_{12}=\frac{e}{\cos\theta}X_{12}=M_H^2.
			\ee

\section{Pure SU(2) vacuum equations}\label{sec4}
			
		We will devote the remainder of the paper to determining the leading order contribution to the vector condensates present for magnetic fields just below $B_{c2}$. As we have shown,(\ref{abrik}) and (\ref{boundary}), this amounts to solving the classical vacuum equations for a pure SU(2) theory, 
		\be
		\mathfrak{\bar D}W=0,\quad
		gw^3_{12}=\frac{g^2}{2}\vert W\vert^2,\label{chrommag}
		\ee
		but with SU(2) broken by the small Higgs condensate. We can emulate this explicit symmetry breaking by introducing a boundary condition
			\be
			\braket{w_{12}^3}=eB_{ext} \neq 0.
			\ee
		This problem was first treated in \cite{Ambjorn:1979xi} using an Abrikosov-like approach, which should be valid for small condensate and linearised equations. A more accurate approach was given in \cite{Olesen:1991df}, where solutions for rectangular lattices in terms of Weierstrass elliptic functions are obtained. Our approach is similar to this, but does cover the case of a hexagonal vortex lattice which is expected to be the true ground state.
		
	\subsection{General solution}

		We rewrite (\ref{chrommag}) as
			\begin{align}
			\bar\partial \ln W&=-ig\bar W^3\label{one},\\
			\partial\ln W^*&=ig W^3\label{two}, \\
			ig(\bar\partial W^3-\partial\bar W^3)&=-g^2\vert W\vert^2,
			\end{align}
		from which we derive:
		\begin{align}
			\Delta \ln\vert W\vert^2+g^2\vert W\vert^2&=0\label{master},\\
			\Delta\arg (W)+\frac{g}{2}(\bar \partial W^3+\partial\bar W^3)&=0\label{argeq}.
		\end{align}
		Choosing the gauge $\partial\cdot W^3=\bar\partial W^3+\partial\bar W^3=0$ makes the second term of the second equation disappear and $\arg(W)$ should be a harmonic function. Naively, one would say that the residual gauge symmetry would allow for \emph{any} such function, but one must keep in mind that the non trivial topology of the vortices gives further restrictions. Taking that into account will allow us to determine the W condensate up to "small" gauge transformations only.
		
		Now we turn to (\ref{master}), which is called a Liouville equation and has solutions of the form \cite{Crowdy}
			\be
			g\vert W\vert=\frac{2\sqrt{2}\vert w\vert}{\vec y(z)^\dag\cdot M\cdot \vec y(z)}\label{liou}.
			\ee
		Here $M$ is a hermitian, positive definite 2-by-2 matrix with $\det(M)=1$ and $\vec y=(y_1,y_2)$ are solutions of the equation
			\be
			y''+\frac{1}{2}E(z)y=0\label{lineq},
			\ee
		with constant Wronskian $w$ and some function $E(z)$, in which the boundary conditions are encoded. 
		
		One can prove
			\be
			\frac{1}{2}E(z)=\frac{d^2}{dz^2} \ln \vert W\vert-(\frac{d}{dz}\ln\vert W\vert)^2 \label{Ez},
			\ee
		while (\ref{one}) can be written as
			\be
			\frac{d}{dz}\ln\vert W\vert=-i(\frac{d}{dz}\arg(W)+\frac{g}{2}\bar W^3)\label{onebis}.
			\ee
		Let us discuss both terms on the right hand side in the gauge $\partial\cdot W^3=\bar\partial W^3+\partial\bar W^3=0$ or, because of (\ref{argeq}), with $\arg(W)$ a sum of a holomorphic and anti-holomorphic function. 
		
		The singular part of the argument of $W$ with anti-vorticity $n$ will in the vicinity of a single vortex be
			\be
			\arg(W)\approx \frac{in}{2}\ln\frac{z}{\bar z},
			\ee
		 in a gauge where $W^3$ is non-singular. We choose $n$ in this way because we know from previous work \cite{Chernodub:2012fi} that the negatively charged $W$ condensates have negative winding, while the positively charged $W^\dag$ have positive vorticity. It will become clear that also in the current approach only for this ansatz solutions can be found. We thus find that the singular part of the right hand side of (\ref{onebis}) is given by:
			\be
			\frac{d}{dz}\ln\vert W\vert=\frac{n}{2z}+\ \mathrm{finite\ part},
			\ee
		such that we find from (\ref{Ez}) for the meromorphic $E(z)$:
			\be
			\frac{1}{2}E(z)=-\frac{2n+n^2}{4z^2}+\frac{\mathrm{residu}}{z}+h(z)\label{Ez2},
			\ee
		with $h(z)$ some undetermined holomorphic function. 
	
		Before proceeding to the solution of (\ref{master}), we can present a slightly different approach to obtaining the form (\ref{liou}). As it is a solution to pure gauge equations of motion, the solution should be locally pure gauge and therefore have the form
			\be
			gW_\mu^a T^a=\left(\begin{array}{cc} \frac{gW_\mu^3}{2}&\frac{gW_\mu^+}{\sqrt{2}}\\\frac{gW_\mu^-}{\sqrt{2}}&-\frac{gW_\mu^3}{2}\end{array}\right)=iU^{-1}\partial_\mu U.
			\ee
		for some $U\in SU(2)$. Like before we make the restriction of this problem to the transversal plane with $\bar W=W^-_1-iW^-_2=0$, and use the conditions (\ref{one}-\ref{two}):
			\begin{align}
			U^{-1}\partial U&=\left(\begin{array}{cc} -\partial\ln \sqrt{W^*}&0\\-\frac{ig}{\sqrt{2}}W&\partial\ln \sqrt{W^*}\end{array}\right)\label{Uunbar},\\
			U^{-1}\bar\partial U&=\left(\begin{array}{cc} \bar\partial\ln \sqrt{W}&-\frac{ig}{\sqrt{2}}W^*\\0&-\bar\partial\ln \sqrt{W}\end{array}\right)\label{Ubar}.
			\end{align}
		The solution of the above system of equations for $U$ gives four independent integration 'constants' $\tilde y_i(z)$,  $i=1,\ldots 4$:
			\be
			U=\left(\begin{array}{cc} \sqrt{W}\tilde y_1^*(\bar z)&\sqrt{W^*}\tilde y_3(z)\\\sqrt{W}\tilde y_2^*(\bar z)&\sqrt{W^*}\tilde y_4(z)\end{array}\right)\label{C11},
			\ee
		for which one essentially uses the triangular character of the coefficient matrix and the diagonal elements. Unitarity gives us then
			\be
			\tilde y_1=\pm \tilde y_4,\quad \tilde y_2=\mp \tilde y_3,\quad \vert \tilde y_1\vert^2+\vert \tilde y_2\vert^2=\frac{1}{\vert W\vert},
			\ee
		With the relative signs related to the freedom in the choice of the square root branch. Now we can calculate with $w(z)=y_1(z)y_2'(z)-y_1'(z)y_2(z)$:
			\begin{align}
			U^{-1}\partial U
			&=\left(\begin{array}{cc} -\partial\ln \sqrt{W^*}&0\\\pm 2Ww^*(\bar z)&\partial\ln \sqrt{W^*}\end{array}\right),\\
			U^{-1}\bar \partial U&=\left(\begin{array}{cc} \partial\ln \sqrt{W}&\pm 2W^*w(z)\\0&-\partial\ln \sqrt{W}\end{array}\right),
			\end{align}
		and find the constant Wronskian condition $\vert w\vert=\frac{1}{\sqrt{8}}$, entirely consistent with the solution (\ref{liou}). The $\tilde y_i$ can be found from the $y_i$  by diagonalising the matrix $M$ and properly normalising and choosing the phase factor to match Eq.(\ref{Uunbar}-\ref{Ubar}). We then have a $SU(2)$ gauge transformation that locally gauges away the Liouville solution (\ref{liou}).
	
\subsection{Lattice solutions}

		Now we will try to put the solutions of the Liouville equation (\ref{liou}) on a 2 dimensional lattice, by demanding double periodicity in the transversal plane such that each lattice cell contains a single vortex. From (\ref{Ez}) we find that then the $E(z)$ are elliptic functions, with the interesting consequence that the 'residu' in (\ref{Ez2}) must vanish for such functions and the holomorphic part must be a constant $h$. It is not hard to to see from (\ref{Ez}-\ref{onebis}) that such a constant can be gauged away by local $U(1)$ gauge transformations. Therefore we can state that a vortex around $\vert z\vert=0$ in a vortex lattice should correspond to:
			\be
			\frac{1}{2}E(z)=-\frac{2n+n^2}{4z^2},\quad z\rightarrow 0\label{Ez2b},
			\ee
		Now the only elliptic function of the form (\ref{Ez2}) for every vortex in the lattice, with only one singular point per lattice cell is proportional to the Weierstrass $\PP$-function:
			\be
			\PP(z)=\sum_{\mathrm{lattice}}\frac{1}{z^2}.
			\ee
		Also no analytic elliptic functions exist, so $\PP(z)$ is indeed unique for our purpose and we can thus rewrite (\ref{Ez2}) as:
			\be
			\frac{1}{2}E(z)=-\frac{2n+n^2}{4}\PP(z)\label{Ez3}.
			\ee
		We will construct lattices of vortices of unit winding $n=1$ and consider only that case from now on.
	
		It is worth noting that the solutions constructed in \cite{Olesen:1991df} over a square lattice correspond to, in our notations:
			\be
			y_1(z)=\frac{2}{\sqrt{\Pp(z/2)}},\quad y_2(z)=\frac{\PP(z/2)}{2\sqrt{\Pp(z/2)}},\label{olesen}
			\ee
		and with some algebra one finds for these functions
			\be
			\frac{y''_i(z)}{y_i(z)}=\frac{1}{8}\left(\frac{3}{2}\left(\frac{\PP''(z/2)}{\Pp(z/2)}\right)^2-\frac{\PP'''(z/2)}{\Pp(z/2)}\right)=\frac{3}{4}\PP(z),
			\ee
		confirming (\ref{Ez3}) with $n=1$.
	
	
		We will now try to solve (\ref{liou}) on a hexagonal lattice 
			\be
			\Pp^2(z)=4\PP^3(z)-\lambda^{-6},\quad \Pp'(z)=6\PP^2(z),\label{P}
			\ee
		assuming the $y_i(z)$ functions of $\PP(z)$:
			\be
			y_i(z)=Y_i(4\PP^3(z)\lambda^6).
			\ee
		This indeed simplifies the equation (\ref{lineq}) to 
			\be
			 \left[ X(1-X)Y''_i+(c-X(1+a+b))Y'_i-ab Y_i\right]\PP=0,\qquad\ \label{hyper}
			\ee
		with 
			\be
			X(z)=4\lambda^6\PP^3(z)\label{X},
			\ee
		and with 
			\be
			a+b=\frac{1}{6},\ ab=-\frac{2n+n^2}{144},\ c=\frac{2}{3},\label{abc}
			\ee
		which is the hypergeometric differential equation. 
		It is known that the equation (\ref{hyper}) has 3 singular points $X=0,1$ and $\infty$, corresponding to the dual lattice points, edge-centers and lattice center points, respectively.
		
			\begin{figure}[h]
				\begin{center}
				\begin{tabular}{cm{0.8cm}c}
					\includegraphics[scale=0.2]{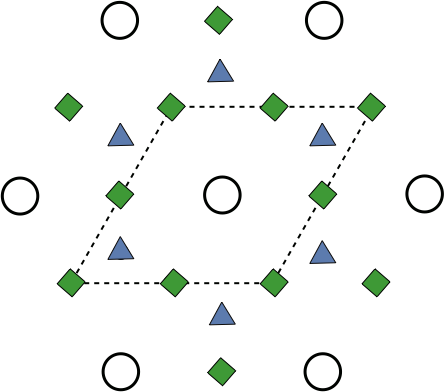}& 
					$\quad\rightarrow$\vspace{25mm}&
					\includegraphics[scale=0.25]{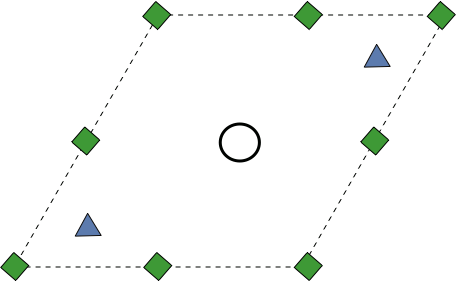}
				\end{tabular}
				\end{center}
				\vskip -5mm
				\caption{The singular points  of the $y_i(z)$ both on the full hexagonal lattice (left) and restricted to just one lattice cell (right):  The vortex-centers (white circles) correspond to $X(z)=\infty$, $X(z)=0$ gives dual lattice points (blue triangles) and $X(z)=1$ mid-points between centers (green squares).}
				\label{fig2}
			\end{figure}
		 
		We will pick two independent solutions such that the first is regular (in the sense that it has a well defined derivative) at $X=0$ and the second at $X=1$ \cite{AS}:
			\begin{align}
			Y_1(X)&=\,_2\mathrm{F}_1(a,b;\frac{2}{3};X),\\
			Y_2(X)&=\lambda\  \,_2\mathrm{F}_1(a,b;\frac{1}{2};1-X),
			\end{align}
		with $\,_2\mathrm{F}_1$ the Gaussian hypergeometric function and $w$ such that
			\be
			\vert Y_2'(X)Y_1(X)-Y_1'(X)Y_2(X)\vert =\vert w\vert\vert\lambda\vert\left\vert\frac{(X/4)^{-2/3}}{12 \sqrt{X-1}}\right\vert,
			\ee 
		which defines the Wronskian for  $y_1$ and $y_2$. To see this, note that  from (\ref{P}) and (\ref{X}) it follows that
			\be
			\frac{dX}{dz}(z)=12\lambda^{-1}(X(z)/4)^{2/3} \sqrt{X(z)-1}. 
			\ee
		One finds for the above choice of $Y$'s:
			\be 
			\vert w\vert \approx 0.25\ \label{wron}.
			\ee
			
		These choices of $Y_i(X)$ and thus of $y_i(z)$ in (\ref{liou}) give a continuous solution over the lattice for any valid choice of the matrix $M$, but extra conditions are necessary to get a smooth solution. To cancel the singular derivative $Y'_1(1)$ in the derivative of $\vert W\vert$ we need to ask
			\be
				(1,\ 0)\cdot M \cdot \vec Y(1)=0,
			\ee
		a condition on the first row of $M$. Similarly the singular $Y'_2(0)$ gives
			\be
				(0,\ 1)\cdot M \cdot \vec Y(0)=0,
			\ee
		a condition on the second row. As $M$ is also required to be hermitic, to have $\det(M)=1$ and be positive definite, we have fixed all its degrees of freedom. The result is then:
			\be
			M=\frac{\left(\begin{array}{cc}
				Y_2(1)Y_2(0)&-Y_1(1)Y_2 (0)\\
				-Y_2(0)Y_1 (1)& Y_1(0)Y_1(1)
				\end{array}\right)}{\sqrt{Y_2(1)Y_2(0)Y_1(1)Y_1(0)-[Y_2(0)Y_1(1)]^2}}\label{M}.
			\ee
			
		Finally,  the scale $\lambda$ is determined by the flux quantization condition 
			\be
			2\pi=\iint g w_{12}^3 dx^2=\mathcal{A}\langle w_{12}^3\rangle,\label{fluxQ}
			\ee
		with $\mathcal{A}$ the surface of one lattice cell given by $\mathcal{A}=2\sqrt{3}\omega_1^2$ if $\omega_1$ is the half-period in the real direction. For the Weierstrass functions given by (\ref{P}) one  can show  \cite{AS} $\lambda\Gamma^3\left(1/3\right)=4\pi\omega_1$ and so $\lambda$ is given by:
			\be
			\frac{\lambda \nu \Gamma^3(1/3)}{2\pi}=\sqrt{\frac{2\pi}{g \langle w^3_{12}\rangle}},\quad \nu=\sqrt{\frac{\sqrt{3}}{2}}.
			\ee
		
		As for the phase, we need some function $\sigma(z)$ such that 
			\be
			\frac{\partial^2}{\partial z^2}arg(W)=-\frac{in}{2}\frac{\partial^2}{\partial z^2}\ln\frac{\sigma^*(\bar z)}{\sigma(z)}=- \frac{in}{2}\PP(z),
			\ee
		which is by definition the Weierstrass $\sigma$-function and then $\arg(W)\equiv -\arg(\sigma)$.
		
		In conclusion, the hexagonal lattice of chromomagnetic anti-vortices with unit winding is:
			\be
			gW(z,\bar z)=\frac{2\sqrt{2}\vert w\vert}{\vert \vec y^\dag(\bar z)\cdot M\cdot \vec y(z)\vert}\frac{\sigma^*(\bar z)}{\vert \sigma(z)\vert},\label{W}
			\ee
			with $y_i(z)=Y_i(4\PP^3(z)\lambda^6)$, $\PP$ and $\sigma$ the Weierstrass $\PP$ and $\sigma$ functions, and with the constant $M$ and $w$ given by (\ref{M}) and (\ref{wron}).
		
		It is remarkable that the profile found in this way is very similar to the one obtained by a more straightforward linearised approach with the ansatz 
			\be
			W\sim e^{-\frac{\langle gw^3_{12}\rangle}{4}\vert z\vert^2}f(\bar z),
			\ee
		used for example in \cite{Ambjorn:1979xi}.
			  
				\begin{figure}[h]
				\begin{center}
					\includegraphics[scale=0.2,clip=false]{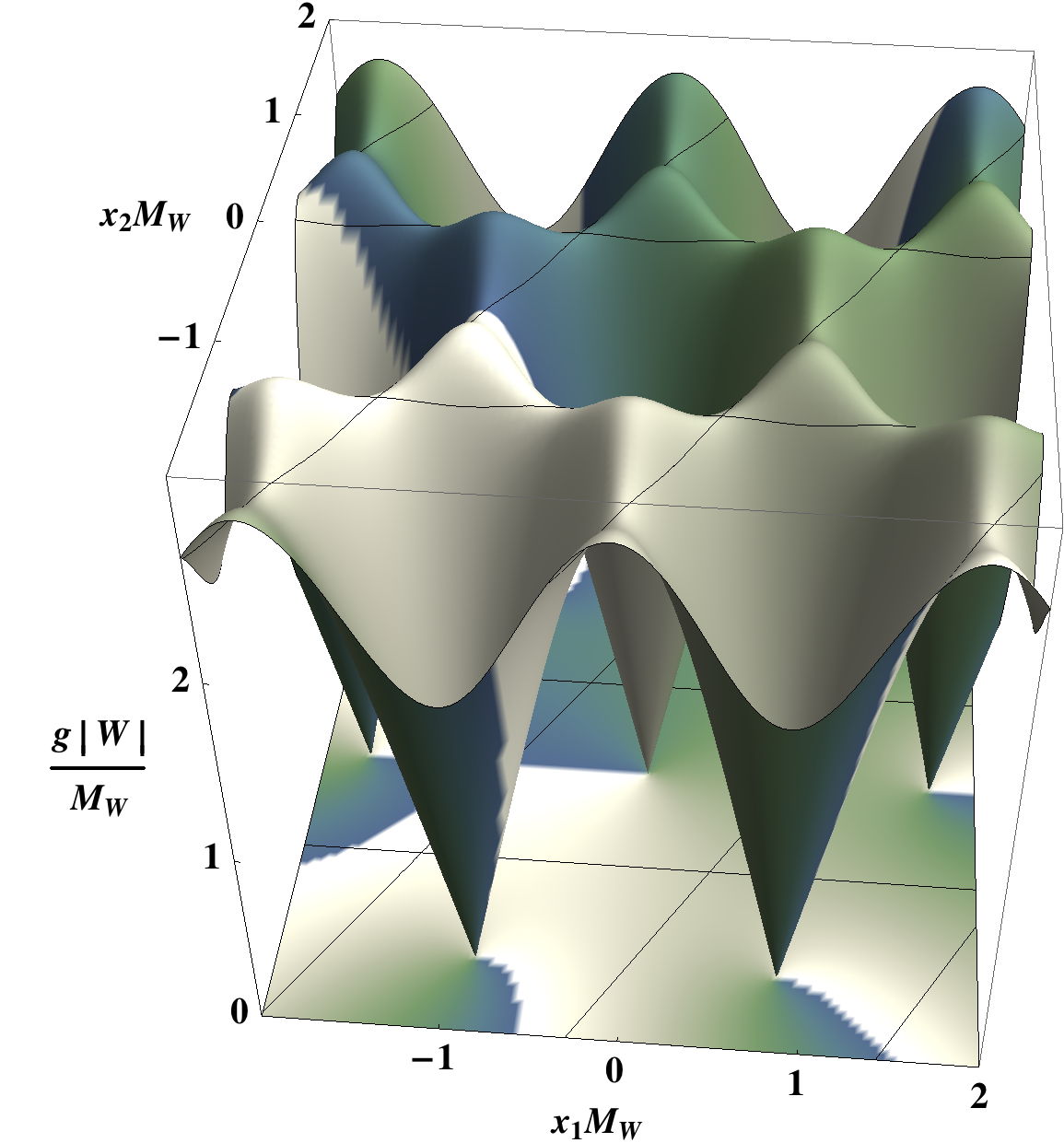}
				\end{center}
				\vskip -5mm
				\caption{The absolute value of the $W$- condensate as function of coordinates on the transversal plane. The dark-to-light fill indicates the rising phase, marking the branch points. The scale is determined by the applied magnetic field, chosen here as $g\langle w_{12}^3\rangle=(5/4)^4M_W^2$.}
				\label{fig3}
				\end{figure}  
			
			
		Our main goal was to obtain a hexagonal structure, but one can prove that for Weierstrass functions obeying 
			\be
			\Pp^2(z)=4\PP^3(z)-\lambda^{-4}\PP(z),\quad \Pp'(z)=6\PP^2(z)-\lambda^{-4},\label{P}
			\ee
		and corresponding to a square lattice, the same hypergeometric equation (\ref{hyper}) as before is found with
			\be
			X=4\PP(z)^2\lambda^4,\quad a=-1/8,\quad b=3/8,\quad c=3/4,
			\ee 
		instead of (\ref{X}) and (\ref{abc}). Repeating the rest of the construction leads to a square lattice solution identical to \cite{Olesen:1991df}, but here written in more complicated algebraic form. 
		
	\subsection{Stability}
		
		We have constructed the lattice solutions in the assumption that SU(2) breaking by a tiny Higgs allowed us to fix $\braket{w_{12}^3}$ to a non-zero value, given by the magnetic flux. This situation changes for vanishing Higgs condensates, beyond $B_{c2}$.
		
		Olesen argued that this lattice, as a solution of the pure gauge equation (\ref{chrommag}), could be stable in a SU(2) invariant theory only if each vortex would have non-trivial topology in the SU(2) group. As $\pi_1(SU(2))=\mathbb{Z}_2$, this corresponds to the existence of \emph{Alice strings} \cite{Volovik:2003fe}. This question can be answered  by calculating the sign of the Wilson loop over one lattice cell, one\footnote{This was shown to me by P.Olesen by means of an explicit calculation in the case of \cite{Olesen:1991df}} can show that the magnetic flux and the condensate contribute compensating signs to the Wilson loop and therefore the vortex does not carry topological charge. The vortices can therefore be gauged away once the symmetry is restored. 
		
		We can recover Olesen's result if we look at the form of $U$ in (\ref{C11}). There we actually constructed the gauge transformation that makes the condensate and the magnetic flux cancel, which is part of the SU(2) symmetry group if no singularities appear. One can easily check that, given the expressions for the $y$ or $\tilde y$ and for the W functions, singularities in (\ref{C11}) could only occur at the zeroes of W. However, the most singular of the $\tilde y_i$ will behave like $(W)^{-\frac{1}{2}}$ there and thus the singularities cancel against the $\sqrt{W}$ factor. Therefore, no vortices and no chromomagnetic flux appear in the symmetry restored phase, and all flux is carried by the hyperflux $X_{12}$. 
		
		Also below $B_{c2}$ a degeneration of the vacuum lattice structure seems to appear in our approximation. The fact that multiple lattice structure solve the pure $SU(2)$ equations is not surprising. It is known that in the Bogomolnyi limit of the electroweak model a similar degeneracy appears. In that case the fourth order terms depending on the Abrikosov parameter \cite{Abrikosov} for the vortex lattice are absent and the energy is a sum of squares. The pure $SU(2)$ energy density has the same property. Rather than a fixed lattice structure, vortices are free as long as they have fixed density given by the flux quantisation condition(\ref{fluxQ}). The vacuum is then a \emph{vortex liquid}.
		
		While in the presence of a small symmetry breaking Higgs condensate the obtained solutions are still good approximations, the liquid property will be lost. All previous calculations of lattices in the superconducting phase between $B_{c1}$ and $B_{c2}$, shown in Fig.\ref{diagram} indicate that then the hexagonal structure should be energetically more advantageous. Therefore, our new solution, while trivial in the symmetry restored phase, is the leading contribution in the presence of a small Higgs condensate.
		
\section{Conclusions}
	
		We have taken a closer look to the phase transition that is reached when increasing the magnetic background even further in the superconducting electroweak vacuum, as depicted in Fig.\ref{diagram}. It is already known that the superconducting phase transition at $B=B_{c1}$ changes the vacuum. This superconducting phase exists in the region of magnetic field strength $B_{c1} < B < B_{c2}$, and has the form of an Abrikosov lattice with a hexagonal structure, in which $W$, $Z$ and Higgs condensates are organised. $B_{c2}$ is reached when the order parameter of the Higgs mechanism, the expectation value of the Higgs field, vanishes. 
		
		We have described the second phase transition at $B=B_{c2}$ using roughly the same approach as when describing the first transition: By considering the condensates on one end of the phase transition to be perturbations of those on the other end. While a similar approach was straightforward when the 'ground state' was the ordinary Higgs vacuum, up to now the exact vacuum in the chromomagnetic vortex state was known for a square lattice only \cite{Olesen:1991df}. 
		
		Furthermore we were able to confirm that qualitatively the phase transition proceeds outside the Bogomolnyi limit in the same way as was found in \cite{Ambjorn:1989bd}, and that the critical field is given by (\ref{bc2}).
		
		The author thanks Maxim Chernodub, Henri Verschelde and Poul Olesen for discussions and suggestions. A first version of this paper was submitted unaware of \cite{Olesen:1991df}, many of the points made already appeared there in a slightly different form. However, it seems that the approaches are sufficiently different to make this work an interesting addition.


\begin{thebibliography}{99} 

	\bibitem{Chernodub:2010qx}
	  M.~N.~Chernodub,
	{\it  ``Superconductivity of QCD vacuum in strong magnetic field,''}
	  Phys.\ Rev.\  {\bf D82}, 085011 (2010).
	  [arXiv:1008.1055 [hep-ph]].
	
	\bibitem{Abrikosov}
	A.A.~Abrikosov, Fundamentals of the Theory of Metals (North Holland, Amsterdam, 1988);
	  A.~A.~Abrikosov,
	{\it ``On the magnetic properties of superconductors of the second group,''}
	  Sov.\ Phys.\ JETP {\bf 5}, 1174-1182 (1957).
	B.~Rosenstein and D.~Li, 
	{\it ``Ginzburg--Landau theory of type II superconductors in magnetic field,''}
	Rev. Mod. Phys. {\bf 82}, 109 (2010).
	
	\bibitem{Ambjorn:1988tm}
	  J.~Ambjorn, P.~Olesen,
	{\it ``On Electroweak Magnetism,''}
	  Nucl.\ Phys.\  {\bf B315}, 606 (1989);
	{\it ``A Magnetic Condensate Solution Of The Classical Electroweak Theory,''}
	  Phys.\ Lett.\  {\bf B218}, 67 (1989).
	
	\bibitem{Chernodub:2011}
	M.N. Chernodub, J. Van Doorsselaere, H. Verschelde, {\it "Electromagnetically superconducting phase of vacuum in strong magnetic fields"}, [arXiv:1111.4401]
	
	\bibitem{Chernodub:2012fi} 
	  M.~N.~Chernodub, J.~Van Doorsselaere and H.~Verschelde, {\it 	Magnetic-field-induced superconductivity and superfluidity of W and Z bosons},  arXiv:1203.5963 [hep-ph].
	
	\bibitem{Linde:1975gx} 
	  A.~D.~Linde, {\it Symmetry Behavior in External Fields},
  	Phys.\ Lett.\ B {\bf 62}, 435 (1976).
  	A.~Salam and J.~A.~Strathdee, {\it Transition Electromagnetic Fields in Particle Physics},  Nucl.\ Phys.\ B {\bf 90}, 203 (1975).

	\bibitem{Grasso:2000wj}
	  D.~Grasso and H.~R.~Rubinstein,
	  {\it Magnetic fields in the early universe,}
	  Phys.\ Rept.\  {\bf 348} (2001) 163
	  [astro-ph/0009061].
	  D.~Comelli, D.~Grasso, M.~Pietroni and A.~Riotto,
	  {\it The Sphaleron in a magnetic field and electroweak baryogenesis,}
	  Phys.\ Lett.\ B {\bf 458}, 304 (1999)
	  [hep-ph/9903227].
	
	\bibitem{Ambjorn:1989bd}J.~Ambjorn, P.~ Olesen {\it A condensate solution of the electroweak theory that interpolates between the broken and the symmetric phase}, Nucl.Phys.B330, 193

	\bibitem{Olesen:1991df} 
	  P.~Olesen,
	  Phys.\ Lett.\ B {\bf 268}, 389 (1991).

	\bibitem{Nambu:1977ag}     
	Y.~Nambu,{\it  String-Like Configurations in the Weinberg-Salam
	                        Theory} Nucl.Phys.B130, 505
	  F.~R.~Klinkhamer and N.~S.~Manton,
	  {\it A Saddle Point Solution in the Weinberg-Salam Theory,}
	  Phys.\ Rev.\ D {\bf 30}, 2212 (1984).
	  T.~Vachaspati,
	  {\it Vortex solutions in the Weinberg-Salam model,}
	  Phys.\ Rev.\ Lett.\  {\bf 68} (1992) 1977
	
	\bibitem{Garriga:1995fv}
	J.~Garriga, X.~ Montes, {\it Stability of Z strings in strong magnetic fields}", Phys.Rev.Lett.75,2268, hep-ph/9505424

	\bibitem{Savvidy:1977as} 
	  G.~K.~Savvidy,
	  "Infrared Instability of the Vacuum State of Gauge Theories and Asymptotic Freedom,''
	  Phys.\ Lett.\ B {\bf 71}, 133 (1977).
		
	\bibitem{Crowdy} D.~Crowdy {\it General Solutions to the 2D Liouville Equation}, Int. J. En. Sci. 35, No. 2, 141, 1997
	
	\bibitem{AS} Abramowitz, M.; Stegun, I.~A., Eds.; {\it Handbook of Mathematical Functions}, 1972.
	
	\bibitem{Ambjorn:1979xi}
	J.~Ambjorn, P.~ Olesen, {\it On the Formation of a Random Color Magnetic Quantum
     	 Liquid in QCD}, Nucl.Phys.B170,60; J.~Ambjorn, P.~ Olesen, {\it A Color Magnetic 		Vortex Condensate in QCD.}, Nucl.Phys.B170,265 

	\bibitem{Volovik:2003fe} 
	  G.~E.~Volovik,
	  {\it The Universe in a helium droplet}
	 Clarendon Press, Oxford (2003).
  	U.~Leonhardt and G.~E.~Volovik,
  	{\it How to create Alice string (half quantum vortex) in a vector Bose-Einstein condensate,}
 	 Pisma Zh.\ Eksp.\ Teor.\ Fiz.\  {\bf 72}, 66 (2000)
  	[JETP Lett.\  {\bf 72}, 46 (2000)]
  	[cond-mat/0003428].
	
\end{thebibliography}
\end{document}